\pdfoutput=1
\documentclass[aps,prd,nofootinbib,twocolumn,citeautoscript,showkeys,superscriptaddress]{revtex4-1} 
\usepackage{amsmath}   
\usepackage{graphicx}  
\usepackage{dcolumn}   
\usepackage{bm}        
\usepackage{amssymb}   
\usepackage{epsfig}
\usepackage{epstopdf}  
\usepackage{subfigure}
\usepackage[utf8]{inputenc}
\usepackage{mdframed}
\usepackage{amsmath}
\usepackage[utf8]{inputenc}
\usepackage{cancel}
\usepackage[normalem]{ulem}
\usepackage{wasysym}
\usepackage[bookmarks=false]{hyperref} 
\hypersetup{pdfstartview=FitH,pdfhighlight=/O,colorlinks=true}
\usepackage{tabu}
\usepackage[toc]{appendix}
\usepackage{color,soul} 
\usepackage{datetime}

\newcommand{\be}{\begin{equation}}
\newcommand{\ee}{\end{equation}}
\newcommand{\tr}[1]{\text{Tr}\left(#1\right)}

\begin{document}

\title{Supersymmetric embedding of antibrane polarization}
\author{Lars Aalsma} 
\email{l.aalsma@uva.nl}
\affiliation{Institute for Theoretical Physics Amsterdam, Delta Institute for Theoretical Physics,
University of Amsterdam, Science Park 904, 1098 XH Amsterdam, The Netherlands}
\affiliation{Institute for Theoretical Physics, KU Leuven,
Celestijnenlaan 200D, B-3001 Leuven, Belgium}
\author{Magnus Tournoy}
\email{magnus.tournoy@kuleuven.be}
\affiliation{Institute for Theoretical Physics Amsterdam, Delta Institute for Theoretical Physics,
University of Amsterdam, Science Park 904, 1098 XH Amsterdam, The Netherlands}
\affiliation{Institute for Theoretical Physics, KU Leuven,
Celestijnenlaan 200D, B-3001 Leuven, Belgium}
\author{Jan Pieter van der Schaar}
\email{j.p.vanderschaar@uva.nl}
\affiliation{Institute for Theoretical Physics Amsterdam, Delta Institute for Theoretical Physics,
University of Amsterdam, Science Park 904, 1098 XH Amsterdam, The Netherlands}
\author{Bert Vercnocke}
\email{bert.vercnocke@kuleuven.be}
\affiliation{Institute for Theoretical Physics, KU Leuven,
Celestijnenlaan 200D, B-3001 Leuven, Belgium}
\date{Wednesday 11$^{\rm th}$ July, 2018}

\begin{abstract}
We study the supersymmetry breaking induced by probe anti-D3-branes at the tip of the Klebanov-Strassler throat geometry. Antibranes inside this geometry polarize and can be described by an NS5-brane wrapping an $S^2$. When the number of antibranes is small compared to the background flux a metastable state exists that breaks supersymmetry. We present a manifestly supersymmetric effective model that realizes the polarized metastable state as a solution, spontaneously breaking the supersymmetry. The supersymmetric model relies crucially on the inclusion of Kaluza-Klein (matrix) degrees of freedom on the $S^2$ and two supersymmetric irrelevant deformations of ${\cal N}=4$ super-Yang-Mills (SYM), describing a large number of supersymmetric D3-branes in the IR. We explicitly identify the massless Goldstino and compute the spectrum of massive fluctuations around the metastable supersymmetry-breaking minimum, finding a Kaluza-Klein tower with masses warped down from the string scale. Below the Kaluza-Klein scale the massive tower can be integrated out and supersymmetry is realized nonlinearly. We comment on the effect of the Kaluza-Klein modes on the effective description of de Sitter vacua in string theory and inflationary model building.
\end{abstract}

\maketitle

\section{Introduction}

Antibranes in a flux background are a key ingredient in many of the constructions of de Sitter vacua in string theory. In the seminal work of Kachru, Kallosh, Linde and Trivedi (KKLT) \cite{Kachru:2003aw}, anti-D3-branes were used to break supersymmetry and to construct metastable de Sitter vacua in string theory. Their work was a huge breakthrough, although it was unclear at the time how the anti-D3-brane contribution could be incorporated in a manifestly supersymmetric manner. Since then, other aspects of the construction have been scrutinized, such as the possible dangerous backreaction of anti-D3-branes on the internal space \cite{McGuirk2009,Bena2009,Blaback:2012nf,Bena:2012ek,Gautason2013,Bena:2014bxa,Blaback:2014tfa,Bena:2014jaa,Danielsson:2014yga,Michel:2014lva,Bena:2015kia,Cohen-Maldonado:2015ssa,Bena:2016fqp,Cohen-Maldonado:2016cjh,Danielsson:2016cit} and antibrane backreaction on four-dimensional moduli \cite{Moritz:2017xto,Moritz:2018sui}. Recently new doubts have even been cast on general aspects of moduli stabilization, before adding antibranes, in flux compactifications \cite{Sethi:2017phn}. It is thus fair to say that various details of the KKLT construction still need to be better understood. For some recent work discussing these and other issues regarding constructions of de Sitter vacua in string theory, we refer the reader to \cite{Brennan2017,Danielsson:2018ztv,Obied2018,Agrawal2018,Dvali2018,Andriot2018,Denef:2018etk,Roupec:2018mbn,Andriot:2018ept,Ghosh:2018fbx,Conlon:2018eyr,Dasgupta:2018rtp,Cicoli:2018kdo,Kachru:2018aqn,Akrami:2018ylq}.

Nevertheless, a point of view that was stressed in \cite{Michel:2014lva,Polchinski:2015bea}, and which we will also assume in this paper, is that knowing the appropriate effective theory at energies (far) below the string scale should be sufficient to capture the relevant physics and provide additional support for the construction. Precisely in this direction there has been much progress over the last few years. As was discussed in \cite{Kallosh:2014wsa,Bergshoeff:2015jxa} (see \cite{McGuirk:2012sb} for an earlier related work), the anti-D3-branes used in KKLT break the supersymmetry of the background, but still preserve supersymmetry nonlinearly, signaling that supersymmetry is broken spontaneously \cite{Volkov:1973ix}. In case of a single anti-D3-brane on top of an orientifold 3-plane, its contribution to the potential can then effectively be captured at low enough energies by a single nilpotent chiral superfield that only contains a Goldstino fermion \cite{Kallosh:2014wsa,Bergshoeff:2015jxa}. Upon removing the orientifold, it was furthermore shown in \cite{Vercnocke:2016fbt,Kallosh:2016aep} that the other worldvolume degrees of freedom can also be captured in terms of additional superfield constraints. These developments have cleared the way for the embedding of inflationary (and de Sitter) models into effective four-dimensional supergravity theories. For a review on this subject, see \cite{Farakos:2017bxs}.

Instead of projecting out degrees of freedom by introducing an orientifold plane, we would like to emphasize that constrained superfields should in general emerge from a supersymmetric theory by integrating out massive degrees of freedom \cite{Komargodski:2009rz}. At low enough energies the massive fields can be integrated out and a single Goldstino remains that can be described by a nilpotent chiral superfield. This yields a low-energy effective action that enjoys non-linear supersymmetry, valid below the cutoff scale set by the masses of the superpartners that are projected out by the constraint \cite{Komargodski:2009rz}. If additional low-energy degrees of freedom exist below the supersymmetry breaking scale they can similarly be described by constrained multiplets \cite{Dudas:2011kt,Ghilencea:2015aph,Dudas:2016eej,DallAgata:2016yof}. In order to understand the leading corrections to any low-energy constrained superfield description it is thus of central importance to identify the supersymmetric origin from which it arises.  

Beyond a single anti-D3-brane, in the case of a small number of anti-D3-branes probing the Klebanov-Strassler (KS) geometry\footnote{To study the physics of anti-D3-branes in warped throats, such as in the KKLT construction, it is standard practice to model the throat by the noncompact KS geometry \cite{Klebanov:2000hb}.} \cite{Klebanov:2000hb}, anti-D3-branes can polarize and settle into a metastable state \cite{Kachru:2002gs}. In \cite{Aalsma:2017ulu} three of us found, by considering a four-dimensional vector multiplet and the (truncated) dynamics of polarization, that the non-linear supersymmetry transformations in the metastable vacuum receive corrections significantly below the scale of supersymmetry breaking. Exactly how to interpret and understand the origin of these corrections remained unanswered and in particular we did not show how the metastable polarized vacuum emerges from a theory invariant under linear supersymmetry. 

In this paper, we will address this question and present an effective supersymmetric model obtained by deforming ${\cal N}=4$ SYM by irrelevant deformations that preserve ${\cal N}=1$ supersymmetry. This model contains a metastable vacuum state very similar to the metastable state of \cite{Kachru:2002gs} and breaks supersymmetry spontaneously by a nonzero F-term. This allows us to explicitly identify a massless Goldstino and clarifies the origin of the corrections to the non-linear supersymmetry transformations that were noticed in \cite{Aalsma:2017ulu}. Importantly, restoring linear supersymmetry requires including the full dynamics on the compact $S^2$ and implies that the standard effective reduction to a single (massive) scalar degree of freedom can be misleading and in fact obscures the appearance of a massless Goldstino in the metastable vacuum. Below the Kaluza-Klein scale one can integrate out the massive tower of states on the $S^2$ and a description in terms of a single nilpotent Goldstino superfield, in addition to the four-dimensional vector multiplet, is a good approximation. 

The rest of this paper is organized as follows. In Sec. \ref{sec:2}, we will review brane polarization of probe anti-D3-branes in the KS background, with a particular emphasis on the spectrum of bosonic fluctuations around the polarized metastable vacuum. We will furthermore argue that it is not possible to restore linear supersymmetry using the probe actions considered in \cite{Kachru:2002gs}. We then continue in Sec. \ref{sec:3} to present a specific linearly supersymmetric model and verify that the metastable vacuum that appears in this model has properties that agree with the metastable state found in \cite{Kachru:2002gs}. Finally, we conclude in Sec. \ref{sec:4} and discuss the implications of our results for metastable de Sitter vacua in string theory and inflationary model building.

\section{Observations with Probe Actions}\label{sec:2}

We will start by giving a short recap of some of the main results of Kachru, Pearson and Verlinde (KPV) \cite{Kachru:2002gs} and highlight the features that will be important later on. In particular, we will show that the effective potential they derived is a consistent truncation that is useful to determine some qualitative features of the metastable polarized state, but that it is not suitable to be used as a four-dimensional effective action, since it does not take all of the dynamics on the $S^2$ into account.

Secondly, neither of the probe actions used by KPV, that is the non-Abelian action of anti-D3-branes and the reduced Abelian NS5-brane action, preserve linear supersymmetry when placed inside the KS background. For this reason the probe actions cannot describe a supersymmetric theory in which the metastable state breaks supersymmetry spontaneously. Let us next provide some more details clarifying these statements.

\subsection{Brane polarization and Kaluza-Klein modes}

\begin{figure}[h]
\includegraphics[scale=1]{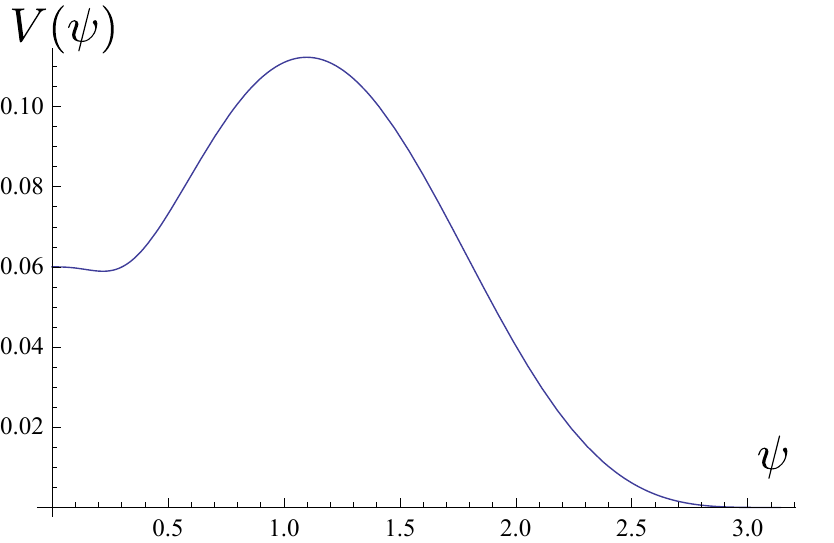}
\caption{The effective potential \eqref{eq:effpot} plotted for $p/M = 0.03$. This potential contains a supersymmetry-breaking metastable minimum when $p/M \ll 1$ and a global supersymmetric minimum at $\psi=\pi$.}
\label{fig:probepot}
\end{figure}

KPV considered creating a nonsupersymmetric solution by adding a small number of anti-D3-branes to the KS geometry \cite{Klebanov:2000hb}. This geometry consists of a long warped throat with a topology $R^{1,3}\times S^3$ at its tip. The throat is supported by putting $M$ units of RR 3-form flux through the $A$-cycle and $K$ units of NSNS 3-form flux through the $B$ cycle of the throat
\be
\frac1{4\pi^2}\int_A F_3 = M~, \qquad \frac1{4\pi^2}\int_B H_3 = -K~.
\ee
At the tip of the throat, the metric is given by
\be
ds^2 = e^{2A_0}\eta_{\mu\nu}dx^\mu dx^\nu + e^{-2A_0}(d\psi^2 + \sin^2\psi ~d\Omega_2^2) ~,
\ee
where $e^{2A_0} \simeq (g_sM)^{-1}\ll 1$ is the warp factor at the tip. Since KPV considered $p>1$ antibranes, the wordvolume description of the anti-D3-branes becomes a non-Abelian gauge theory with gauge group $U(p)$, that contains noncommuting matrix degrees of freedom. This allows for a version of the Myers effect \cite{Myers:1999ps} where the anti-D3-branes polarize to collectively form a spherical 5-brane configuration that has a description for $p \gg 1$ in terms of an NS5-brane\footnote{Formally, the NS5-brane action is strongly coupled as it is derived by S-dualizing the D5-brane action. Nevertheless, KPV found reasonable agreement between the NS5-brane perspective and the non-Abelian anti-D3-brane perspective. } wrapping an $S^2$ of  the $S^3$. Using the NS5-brane perspective, KPV derived an effective potential of the dynamics of the NS5-brane in terms of the azimuthal angle $\psi$ on the $S^3$, given by
\be \label{eq:effpot}
V(\psi) = \sqrt{Q(\psi)^2 + \frac{M^2}{\pi^2}b_0^4\sin^4\psi }  -Q(\psi)  ~,
\ee
with $b_0^2\approx 0.93$. The effective 3-brane charge $Q(\psi)$ is given by
\be
Q(\psi) = \frac{M}{\pi}\left(\psi-\frac12\sin(2\psi)\right) - p~.
\ee
In addition to the nonsupersymmetric metastable minimum present for $p/M\lesssim 0.08$ at $\psi_{\rm min} = \frac{2p\pi}{Mb_0^4} $, this theory also has a global supersymmetric minimum at $\psi=\pi$ where all antibranes have annihilated and linear supersymmetry is restored; see Fig \ref{fig:probepot}.

KPV also described the metastable minimum from the perspective of the non-Abelian worldvolume gauge theory of the anti-D3-branes. In that case, the potential expanded around the nonsupersymmetric pole $(\psi=0)$ is given as a function of three bosonic $p\times p$ dimensional matrix degrees of freedom $\phi^{i=1,2,3}$
\be \label{eq:KPVpot}
V(\phi^i) = 2p + \frac{i}{3}\kappa ~\epsilon_{ijk}\tr{[\phi^i,\phi^j]\phi^k} - \frac \lambda4\tr{[\phi^i,\phi^j]^2} ~.
\ee
We will first discuss the features of this potential and after that match its parameters to the effective potential \eqref{eq:effpot}. Note that for now we have ignored the fermionic degrees of freedom that should in principle also be there. Once we understand the linearly supersymmetric embedding of this model the fermionic sector can straightforwardly be included. 

The critical points of \eqref{eq:KPVpot} are given by
\begin{align} \label{eq:su(2)com}
[\phi^i,\phi^j] &= \frac{\kappa}{\lambda}i\epsilon^{ijk} \phi_k ~, 
\end{align}
which correspond to the commutation relations of the generators of $SU(2)$ upon rescaling $\phi^i = \frac \kappa\lambda J^i$. We see that any $SU(2)$ representation extremizes the potential, but two representations are worth mentioning specifically. The trivial representation (which has commuting matrices) corresponds to parallel anti-D3-branes and has the highest vacuum energy. The configuration of lowest energy is given by the $p$-dimensional irreducible representation, corresponding to the minimum where the anti-D3-branes have polarized \cite{Kachru:2002gs}. The vacuum energy in the metastable minimum, after reinstating the correct units, is given by
\be \label{eq:vacuumenergy}
V_{\rm min} = T_{D3}e^{4A_0}p\left(2-\frac{\kappa^4}{24\lambda^3}(p^2-1)\right)  ~,
\ee
Here, $T_{D3}$ is the tension of an (anti-)D3-brane. The radius of the $S^2$ in the minimum is given by
\be \label{eq:S2radius}
R^2_{S^2} = \frac{e^{-2A_0}}{p}\tr{\phi^i\phi_i}\ell_s^2 = e^{-2A_0}\frac{\kappa^2}{4\lambda^2}(p^2 -1)\ell_s^2 ~,
\ee
where $\ell_s$ is the string length.

Comparing \eqref{eq:vacuumenergy} and \eqref{eq:S2radius} to the equivalent quantities from the Abelian NS5-brane perspective, we find
\be
\kappa^2 = \frac{M^2}{p^2\pi^2}(p^2-1) ~, \quad \lambda = \frac{M^2b_0^4}{4p^2\pi^2}(p^2-1) ~.
\ee
Expanding the matrices around the metastable vacuum in fluctuations as 
\be
\phi^i = \frac \kappa\lambda J^i + \varphi^i ~,
\ee
one finds a tower of fields that are labeled by $l=1,\dots,p-1$.\footnote{The $l=0$ modes are gauge redundancies and should be omitted.} Details on how to explicitly diagonalize the mass matrix can be found for example in \cite{Andrews:2006aw}.

After canonically normalizing the scalar fields, the masses of the different states in the metastable minimum are given by
\be \label{eq:KKmass}
m^2_{\varphi} = \frac{\kappa^2}{\lambda}l(l+1)e^{4A_0}m_s^2 = \frac{4}{b_0^4}l(l+1) e^{4A_0}m_s^2 ~,
\ee
with a $2(2l+1)$ multiplicity for the eigenvalues of the mass matrix for each $l$ \cite{Jatkar:2001uh}. Notice that the masses are warped down from the string scale $m_s$.

From the Abelian NS5-brane perspective we can similarly describe fluctuations around the metastable vacuum by performing a Kaluza-Klein reduction of the NS5-brane action on the $S^2$. One then also finds a tower of states \cite{Bachas:2000ik} that can be matched with \eqref{eq:KKmass}. This emphasizes that the effective potential derived by KPV should be understood as a (bosonic) truncation of the full NS5-brane theory that keeps only one bosonic degree of freedom. This is the azimuthal angle $\psi$, whose mass around the metastable minimum is given by \cite{Aalsma:2017ulu}
\be \label{eq:psimass}
m_\psi^2=\frac{8}{b_0^4}e^{4A_0}m_s^2 ~.
\ee
This mass matches the one calculated from the non-Abelian perspective for $l=1$, see \eqref{eq:KKmass}. We want to stress once more that this degree of freedom is in fact part of a Kaluza-Klein tower of states, which will turn out to be important when we introduce our effective supersymmetric description later on. The fact that this mass does not explicitly feature the radius of the $S^2$, which one might have expected, is an artifact of the fact that the location of the metastable minimum in a small $\psi$ expansion depends on $p/M$, just as the radius of the $S^2$. The overall $p/M$ dependence therefore drops out.

We thus come to the conclusion that the Kaluza-Klein scale in the four-dimensional effective theory is set by the mass of this fluctuation
\be \label{KKpsi}
E_{\rm KK} = m_\psi ~,
\ee
which is warped down from the string scale. Thus, strictly speaking the effective theory in the metastable minimum can only be considered four-dimensional at energy scales $E \ll m_\psi \ll m_s$. Also note that the supersymmetry breaking scale, set by the value of the KPV potential at the metastable minimum, is of the same order as $m_\psi$. This strongly suggests that the restoration of linear supersymmetry hinges on the inclusion of all the Kaluza-Klein modes on the $S^2$, of which the $\psi$ field is just one particular component. 

\subsection{Probes cannot restore supersymmetry}

Having identified the bosonic spectrum in the metastable vacuum, we now turn to discuss supersymmetry. In particular, if the metastable state breaks supersymmetry spontaneously at low energy, it should be possible to find a linearly supersymmetric field theory in which the metastable state is a solution. However, we will now show that the probe actions used by KPV are not linearly supersymmetric when placed inside the KS background and therefore cannot be used to describe spontaneous supersymmetry breaking, at least not within a four-dimensional low-energy effective action.

To illustrate this point, let us consider a single probe anti-D3-brane. The worldvolume theory of an anti-D3-brane (in flat space) has 32 supersymmetries, 16 of which are linearly realized and the other 16 nonlinearly; see for example \cite{Bergshoeff:2013pia}. The supersymmetries that are preserved by the antibrane are solutions to\footnote{See \cite{Simon:2011rw} for a nice review on supersymmetric brane actions.}
\be
(1-\Gamma_{\overline{\rm D3}})\varepsilon = 0 ~,
\ee
where $\Gamma_{\overline{\rm D3}}$ is the $\kappa$-symmetry projector of an anti-D3-brane. On the other hand, because the background we are interested in contains an O3 orientifold projection, the supersymmetries that are preserved by the background are solutions to \cite{Kallosh:2014wsa}
\be
(1-\Gamma_{\rm O3})\varepsilon = 0 ~,
\ee
where $\Gamma_{\rm O3}$ is the action of an O3 orientifold projection. For an anti-D3-brane probing the KS background, $\Gamma_{\overline{\rm D3}} = -\Gamma_{\rm O3}$, which shows that the linear supersymmetries on the anti-D3-brane are projected out by the orientifold and only the non-linear supersymmetries survive. As a result, all degrees of freedom on the antibrane transform in the standard non-linear manner under supersymmetry \cite{Vercnocke:2016fbt,Kallosh:2016aep}, but there is no chance of restoring linear supersymmetry as the linear supersymmetries are projected out.

One might expect that the situation is better from the perspective of the NS5-brane, because the effective potential \eqref{eq:effpot} connects the supersymmetry breaking state to the global supersymmetric vacuum. In this picture however, the supersymmetries preserved by the NS5 brane are solutions to
\be
(1-\Gamma_{\rm NS5}(\psi))\varepsilon = 0 ~.
\ee
The precise form of the four-dimensional reduced $\kappa$-symmetry projector $\Gamma_{\rm NS5}(\psi)$ can be found in ({\color{red} 3.4}) of \cite{Aalsma:2017ulu}. One important aspect is that at the poles of the $S^3$ it is given by
\be
\Gamma_{\rm NS5} = 
\begin{cases}
\Gamma_{\rm \overline{D3}} \quad (\psi=0)~, \\
\Gamma_{\rm D3} \quad (\psi=\pi) ~,
\end{cases}
\ee
but away from the poles it does not align nicely with the $\kappa$-symmetry projector of an (anti-)D3-brane. So in general, away from the poles, the (reduced) NS5-brane projector also breaks supersymmetry explicitly, even though it does interpolate between the D3-brane and anti-D3 brane projector. We conclude that if a supersymmetric theory exists in which the metastable state is a solution it cannot be described by one of the KPV four-dimensional probe actions. The action for probe anti-D3-branes only preserves non-linear supersymmetry and has no connection to the linear regime. The NS5-brane action on the other hand is linearly supersymmetric at $\psi=\pi$, but away from this pole breaks supersymmetry explicitly. We will argue that both probe descriptions do not contain enough degrees of freedom to allow for the restoration of supersymmetry. To restore linear supersymmetry we will be forced to include additional (massive) degrees of freedom on the $S^2$, which are clearly absent in the reduced four-dimensional KPV probe descriptions. 

Nevertheless, at the supersymmetric pole the physics is that of $(M-p) \gg 1$ D3-branes, and we will use a supersymmetric expansion around that theory to get information on the specific deformations that are needed. In the next Sec., we will go beyond the level of the probe action and propose a supersymmetric model of $(M-p)$ D3-branes in the KS background, obtained by introducing additional degrees of freedom from the $S^2$ and specific irrelevant deformations of ${\cal N}=4$ SYM. Adding these degrees of freedom and deformations the supersymmetric model then indeed features a metastable vacuum state in which supersymmetry is spontaneously broken.

\section{Supersymmetric completion of the metastable state}\label{sec:3}

In the previous Sec. we argued that neither of the probe actions used by KPV preserve linear supersymmetry when placed inside the KS background. So if we think that anti-D3-brane polarization can nevertheless be embedded in a supersymmetric effective field theory model, how are we going to identify that theory? 

Of course, any D-brane state is a solution of superstring theory, so it might be the case that an explicit description in terms of spontaneous breaking of supersymmetry is only possible in the full ten-dimensional superstring theory. This is a possibility, but one that is contrary to expectations in this particular KPV setup. Because the supersymmetry breaking scale is warped down from the string scale \cite{McGuirk:2012sb} one would expect that the metastable state can be embedded in a lower-dimensional effective supersymmetric theory. In particular, if backreaction of anti-D3-branes does not destroy the metastable state and a well-defined supergravity solution exists \cite{Cohen-Maldonado:2015ssa}, it must be realizable as a state in a supersymmetric field theory that is (holographically) dual to the supergravity solution \cite{DeWolfe:2008zy}.

For example, before adding anti-D3-branes to the KS geometry, the holographic dual of KS is given by a nonconformal ${\cal N}=1$ cascading gauge theory \cite{Klebanov:2000hb}. The effect of antibranes can then be included by adding a particular nonsupersymmetric perturbation to the cascading gauge theory \cite{DeWolfe:2008zy} and it was shown in subsequent work \cite{Bertolini:2015hua,Krishnan:2018udc} that the resulting gauge theory indeed contains a massless fermion, as expected if the supersymmetry breaking was spontaneous. Although this is suggestive, in these approaches the perturbations describing anti-D3-branes still explicitly break supersymmetry. Our goal here is to instead present an effective fully supersymmetric model in which the polarized antibrane state appears as a metastable nonsupersymmetric solution.

Indications for how to construct such a model can be obtained from the works \cite{Bena:2014jaa,Bena:2015kia,Bena:2016fqp} (for related work, see \cite{Hartnett:2015oda}). Here, the authors argued that, close to the nonsupersymmetric $\psi=0$ pole, anti-D3-branes source an $AdS_5\times S^5$ throat perturbed by flux that is dual to relevant deformations of ${\cal N}=4$ SYM that break all supersymmetry. Hence, the dual gauge theory that describes these antibranes is a nonsupersymmetric version of the ${\cal N}=1^\star$ theory of Polchinski and Strassler \cite{Polchinski:2000uf}. Obviously, because we are interested in finding a supersymmetric starting point, we will not try to identify the polarized state in a theory obtained as an expansion around the nonsupersymmetric ($\psi=0$) pole, but instead we will start from the supersymmetric ($\psi=\pi$) pole. Here, the physics should be that of $(M-p)$ D3-branes that source a supersymmetric $AdS_5\times S^5$ throat dual to ${\cal N}=4$ SYM. To describe polarization, we suggest that one should add irrelevant deformations to ${\cal N}=4$ SYM that have the effect of gluing the $AdS_5\times S^5$ throat back to the Klebanov-Strassler region \cite{Evans:2001zn}. In fact, an expansion of the effective potential \eqref{eq:effpot} around the supersymmetric pole of the $S^3$ reveals that, as anticipated, the first term in the effective potential is $\psi^4$ (which corresponds to the SYM term in the worldvolume gauge theory) and the leading corrections are irrelevant operators (in four dimensions) of mass dimension $6$ and $7$.

We will provide evidence below that a manifestly supersymmetric model that includes matrix degrees of freedom on the $S^2$ and irrelevant operators of mass dimension $6$ and $7$ indeed features a nonsupersymmetric metastable vacuum state with the expected properties. Obviously, this model is fine-tuned in the sense that all other irrelevant operators should be suppressed; that is, it crucially relies on the details of the UV embedding. Nevertheless, if we extrapolate this model away from the supersymmetric pole, we discover a metastable vacuum that has the same properties as the metastable state found by KPV. It breaks supersymmetry spontaneously by a nonzero F-term and as a consequence features a massless Goldstino. This fermion is only massless at the metastable minimum, as appropriate for a Goldstone fermion, and therefore provides an additional low-energy degree of freedom on top of the gaugino residing in the vector multiplet. 

\subsection{The model}

The field content of a stack of $N=(M-p)$ D3-branes in four dimensions is given by three matrix-valued chiral multiplets $\Phi^{i=1,2,3}$ and a matrix-valued vector multiplet. We will ignore the vector multiplet and just focus on the three chiral multiplets, which is sufficient for our purposes. The supersymmetric model we propose to describe the metastable state is defined by the following superpotential
\begin{align} \label{eq:irrelevantsuper}
W(\Phi^i) &= \frac{a}{3!} \epsilon_{ijk}\tr{[\Phi^i,\Phi^j]\Phi^k} \nonumber \\
&+ i\frac {b}{5}\epsilon_{ijk}\tr{[\Phi^i,\Phi^j][\Phi^k,\Phi^l]\Phi_l  } \nonumber \\
&+ \frac{c}{6}\epsilon_{ijk}\epsilon_{lmn}\tr{[\Phi^i,\Phi^j][\Phi^l,\Phi^m][\Phi^k,\Phi^n]  } ~,
\end{align}
and a canonical K\"ahler potential. The first line of \eqref{eq:irrelevantsuper} is the ${\cal N}=4$ SYM term and the second and third lines are the irrelevant deformations, that break supersymmetry from ${\cal N}=4\to1$. Locally, the metric on field space can be approximated by the flat metric $\delta_{ij}${\footnote{We believe our results can easily be generalized beyond this approximation, but it will be sufficient for our purposes here.}} and we take all coefficients to be real.

The scalar potential is given by
\be \label{eq:scalpot}
V_S = \tr{\frac{\partial W}{\partial\Phi^i}\frac{\partial \overline W}{\partial \overline{\Phi}_i} }~.
\ee
To find its critical points, we take the following ansatz for the commutation relations of the scalar fields
\be \label{eq:su(2)}
[\phi^i,\phi^j] = i\epsilon^{ijk}\phi_k ~,
\ee
which are the commutation relations of the generators of $SU(2)$. Just as before, the trivial representation corresponds to parallel D-branes, corresponding to a vacuum with vanishing energy that makes it supersymmetric. In addition, there can be two other types of vacua that obey \eqref{eq:su(2)}, depending on the choice of parameters. They are given by
\begin{align}
\text{Type I:}&\quad a + 2b - 4c = 0 ~, \\ 
\text{Type II:}&\quad a + 4b - 10c = 0 ~. \label{eq:typeIIvac}
\end{align}
The vacuum energy in these two vacua after reinstating units is given by
\begin{align}
\text{Type I:} &\quad V_{\rm min} = 0 ~, \\
\text{Type II:}&\quad V_{\rm min} = T_{D3}e^{4A_0}\left(\frac{N}{100}(N^2-1)(3a +2b)^2 \right) ~.
\end{align}
We see that the type I vacuum has vanishing vacuum energy (and vanishing F-terms) and therefore corresponds to a supersymmetric state, similar to the supersymmetric polarized states in the ${\cal N}=1^\star$ theory \cite{Polchinski:2000uf}. The type II vacuum, however, has nonvanishing F-terms, and positive vacuum energy and it necessarily breaks supersymmetry spontaneously. Upon comparing the vacuum energy with \eqref{eq:vacuumenergy}, we find
\be \label{eq:vacenerg}
a = -\frac23b + \cal{O}(\epsilon) ~,
\ee
where $\epsilon = \sqrt{p/M^3} \ll 1$ is a small parameter. By expanding in fluctuations around the type II vacuum as $\phi^i=J^i + \varphi^i$, we again find a tower of scalar fields. The mass of the lightest fluctuation is given by
\be
m^2_{\varphi} \simeq \frac{24}{5}\left(a\epsilon - \frac43 \epsilon^2 \right)e^{4A_0}m_s^2 ~,
\ee
where we used \eqref{eq:vacenerg}. This result matches \eqref{eq:psimass} when we identify
\be
a =  \frac 5{3b_0^4\epsilon} + {\cal O}(\epsilon) ~.
\ee
Thus, all fluctuations have a positive mass squared when $p/M^3\lesssim 1$. Notice that this condition is weaker than, but consistent with, the condition $p/M \lesssim 0.08$ derived in \cite{Kachru:2002gs} for a metastable minimum to exist. We conclude that the superpotential \eqref{eq:irrelevantsuper} reproduces the main features of the effective potential derived by KPV. As already mentioned, we should stress that since this crucially relies on the inclusion of two (supersymmetric) irrelevant corrections, the model depends sensitively on the UV (string) theory embedding.  

In addition, because the metastable vacuum breaks supersymmetry spontaneously it should have a massless fermion in its spectrum corresponding to the Goldstino of supersymmetry breaking. By computing the determinant\footnote{Because we are working with matrix-valued fields, the fermion mass matrix has four indices. To compute the determinant this tensor first needs to be decomposed to obtain a regular $3N^2\times 3N^2$ matrix of which we can calculate the determinant.} of the fermionic mass matrix
\be
\det(M_F^{ij}) = \det\left(\frac{\partial^2 W}{\partial\Phi_i \partial \Phi_j} \right) ~,
\ee
we indeed find that this vanishes, exactly when \eqref{eq:typeIIvac} is satisfied. 

Obviously, one should also be able to identify this Goldstino from the perspective of an Abelian NS5-brane description expanded around the supersymmetric pole. There it should correspond to a fermionic zero mode on the $S^2$ threaded by magnetic flux (the presence of flux twisting the $S^2$ is crucial to allow for a zero mode \cite{Andrews:2006aw}), but since the degrees of freedom are organized differently in the Abelian NS5-brane perspective identifying the Goldstino is not straightforward and it would be of interest to confirm its presence. Whether such a fermionic zero mode should also be present in the non-Abelian anti-D3-brane description expanded around the nonsupersymmetric pole, as was considered in \cite{McGuirk:2012sb}, is not obvious \emph{a priori}, but it should clearly not be associated with the gaugino. In the absence of a supersymmetric embedding the appearance of a Goldstone fermion should be expected to be difficult at best. 

To summarize, the proposed model allows for a supersymmetric description of brane polarization. It describes both a supersymmetric vacuum corresponding to $N$ parallel D3-branes and a metastable vacuum. The metastable vacuum breaks supersymmetry spontaneously and its spectrum of fluctuations contains the four-dimensional vector multiplet, a massless Goldstino and a massive tower of Kaluza-Klein states that, when included, allow for full restoration of supersymmetry. The supersymmetry breaking scale $\sqrt{f}$ is thus of the order of the Kaluza-Klein scale
\be
\sqrt{f} \simeq E_{\rm KK} = m_{\varphi} ~,
\ee
as can be seen by comparing the potential energy and the mass of the lightest fluctuation. As a consequence, below the supersymmetry breaking scale one can integrate out the massive fields such that a constrained superfield description in terms of a single nilpotent superfield that contains the Goldstino is a good approximation for the dynamics on the $S^2$ and supersymmetry is realized nonlinearly. At or above this scale, this description will be modified and one has to take the Kaluza-Klein modes on the $S^2$ into account. Because the Kaluza-Klein scale is warped down from the string scale, as noted in Sec. \ref{sec:2}, the energy at which one needs to include the Kaluza-Klein modes can be very low. Depending on the details realizing a hierarchy between the energy scale one is probing and the warped down Kaluza-Klein scale might therefore be difficult, but not impossible.

\section{Conclusions}\label{sec:4}

In this paper, we have studied the supersymmetry breaking of anti-D3-branes placed inside the KS geometry. Via brane polarization, the antibranes settle into a supersymmetry-breaking metastable state, when the number of antibranes is sufficiently small compared to the background flux \cite{Kachru:2002gs}. To identify this breaking as spontaneous we set out to identify a supersymmetric field theory model featuring a metastable vacuum state. Our proposal for an effective supersymmetric field theory model indeed exhibits such a metastable supersymmetry-breaking vacuum with all the expected properties. Besides the fact that we ignored gravity by working in the noncompact KS geometry (effectively sending $M_{\rm Pl} \rightarrow \infty$) , this construction is also fine-tuned; that is, the details rely sensitively on the specific UV embedding due to the necessary introduction of irrelevant operators. Keeping those limitations in mind the qualitative features of the metastable solution nicely agree with the polarized state in the KPV model. Expanding around the metastable solution we observed that the restoration of supersymmetry crucially relies on taking into account a tower of Kaluza-Klein modes, with masses warped down from the string scale, explaining why it is impossible for the truncated probe actions to restore supersymmetry. 

Our results have a number of consequences. First of all, the crucial importance of Kaluza-Klein, or equivalently matrix model modes on the $S^2$ confirms that the KPV effective potential should only be thought of as a limited (albeit consistent) truncation of a more complete description. In other words, the additional degrees of freedom related to the $S^2$ only decouple at the poles and should in general be included in the four-dimensional effective theory. This should impact models derived from KPV's effective potential, such as the inflationary model recently considered in \cite{Gautason:2016cyp,DiazDorronsoro:2017qre}, where the effective potential \eqref{eq:effpot} was used for large field inflation in the regime $p/M \gg 1$. Our results clearly suggest that it is inconsistent to rely on just the scalar mode $\psi$ for the effective dynamics on the $S^2$ and one should include (a subset of) a Kaluza-Klein tower of states. As a result one would expect the results reported in \cite{Gautason:2016cyp,DiazDorronsoro:2017qre} to be affected. We note that an inflationary model that is described from the perspective of scalar matrix degrees of freedom, albeit in a different context, already has been studied in \cite{Ashoorioon2009}. In light of the results obtained here it might be of interest to revisit some of these approaches. 

Secondly, the decay rate to tunnel from the metastable state to the global supersymmetric vacuum was calculated in \cite{Kachru:2002gs} by making use of the effective potential \eqref{eq:effpot}. However, in the presence of additional degrees of freedom this tunneling rate will likely be modified. For example, it is well known that a coupling between a quantum mechanical system and its environment can lead to a significant suppression of the tunneling rate \cite{Caldeira:1981rx,Caldeira:1982uj}. Furthermore, additional degrees of freedom can also open up new decay channels; see for example \cite{deAlwis:2013gka} and references therein. If and how these effects modify the lifetime of the metastable vacuum is a question that we hope to return to in future work.

Our results also shed light on how the anti-D3-brane uplift procedure in KKLT might be captured by an effective supersymmetric theory. To be specific, we claim that the constrained superfield description proposed in \cite{Kallosh:2014wsa,Bergshoeff:2015jxa}, with just a single nilpotent superfield containing the Goldstino, is certainly a valid description below the warped down Kaluza-Klein scale. However, to understand this in terms of spontaneous supersymmetry breaking and identify leading corrections one needs to identify the supersymmetric completion of this nonsupersymmetric phase. This requires the introduction of additional degrees of freedom at the warped down Kaluza-Klein scale. The effective field theory model that we introduced to describe the (matrix) dynamics on the compact $S^2$ is a good candidate for a linearly supersymmetric theory in which the polarized state appears as a nonsupersymmetric solution. This model does not contain degrees of freedom at the string or Planck scale, so it can be studied as a supersymmetric low-energy effective field theory, but it does depend sensitively on the specific UV (string theory) embedding. 

Finally we would like to comment on how our results are related to the debate in the literature regarding the existence of the KPV metastable state after taking into account the full backreaction of anti-D3-branes. As mentioned, the nonexistence of a fully backreacted supergravity solution would imply that supersymmetry breaking in the holographically dual gauge theory description is necessarily explicit. Since the effective supersymmetric field theory model that we introduced nicely allows for the appearance of a broken supersymmetry phase with the expected properties to relate it to antibrane polarization, this suggests that a fully backreacted supergravity solution should exist, in line with results of \cite{Cohen-Maldonado:2015ssa}, assuming such a holographic duality.

However, even if the supersymmetric model constructed is holographically dual to this fully backreacted supergravity solution, it is clearly not UV complete and requires a specific UV embedding. Whether or not such an embedding is possible in string theory will determine its ultimate fate. So with these results we can certainly not rule out the possibility that the KPV metastable state and therefore the KKLT de Sitter vacua, after including gravity,  all belong to the swampland, as was recently conjectured \cite{Brennan2017,Danielsson:2018ztv,Obied2018,Agrawal2018}. As a consequence, understanding the UV completion of the proposed model is an important direction for further research.

\section*{Acknowledgments}

We would like to thank Riccardo Argurio, Iosif Bena, Johan Bl\aa b\"ack, Fotis Farakos, Guilherme Pimentel, Thomas van Riet and Gary Shiu for useful discussions.  This work is part of the Delta Institute for Theoretical Physics consortium, a program of the Netherlands Organisation for Scientific Research (NWO) that is funded by the Dutch Ministry of Education, Culture and Science (OCW). 
L.A. gratefully acknowledges KU Leuven for its hospitality, while part of this work was completed and financial support during his stay. Similarly, M.T. would also like to thank the University of Amsterdam for its hospitality, while part of this work was performed and is grateful to the Delta ITP for financing his stay. 
The work of L.A. and J.P.v.d.S. is also supported by the research program of the Foundation for Fundamental Research on Matter (FOM), which is part of the Netherlands Organization for Scientific Research (NWO).  
The work of M.T. is supported by the FWO Odysseus Grant No. G.0.E52.14N. 
B.V. is supported by the European Research Council Grant No.\ ERC-2013-CoG 616732 HoloQosmos, the KU Leuven C1 Grant No. ZKD1118 C16/16/005 and the COST actions CA16104 \emph{GWVerse} and  MP1210 The String Theory Universe.

\end{document}